\documentclass[12pt]{iopart}

\usepackage{graphicx}% Include figure files
\usepackage{dcolumn}% Align table columns on decimal point
\usepackage{bm}
\usepackage[T1]{fontenc}
\usepackage[latin1]{inputenc}
\usepackage[english]{babel}
\usepackage{color}
\usepackage{pgf}
\usepackage{multimedia}
\usepackage{mathptmx}
\usepackage{mathrsfs}
\usepackage{amsfonts}
\usepackage{iopams}
\usepackage{selinput}
\usepackage{ulem}

\begin{document}
%\preprint{1p0mt}

%\newcommand{\Tc}{$T_c$} % \cong$}
%\newcommand{\units}{$\mu \text{W}/\text{K}^2\text{cm}$}
%\newcommand{\p}[1]{\left( #1 \right)}
%\newcommand{\Dd}[2]{\frac{\text{d} #1}{\text{d}#2}}

%%%%%%%%%%%%%%%%%%%%%%%%%%%% TITLE

\title[On the search for the chiral anomaly in Weyl semimetals]{
On the search for the chiral anomaly in Weyl semimetals: The negative longitudinal magnetoresistance
}

%%%%%%%%%%%%%%%%%%%%%%%%%%%% AUTHORS

\author{R.~D.~dos~Reis, M.~O.~Ajeesh, N.~Kumar, F.~Arnold, C.~Shekhar, M.~Naumann, M.~Schmidt, M.~Nicklas, and E.~Hassinger}
\address{Max Planck Institute for Chemical Physics of Solids,  N\"{o}thnitzer Str.\ 40, 01187 Dresden, Germany}

\ead{nicklas@cpfs.mpg.de \textrm{and} elena.hassinger@cpfs.mpg.de}

{\hspace{17mm}\tiny\scriptsize{\today}}

%%%%%%%%%%%%%%%%%%%%%%%%%%%% ABSTRACT

\begin{abstract}
Recently, the existence of massless chiral (Weyl) fermions has been
postulated in a class of semi-metals with a non-trivial energy dispersion.
These materials are now commonly dubbed Weyl semi-metals. One predicted property of Weyl fermions is the chiral or Adler-Bell-Jackiw anomaly, a chirality imbalance in the presence of parallel magnetic and electric fields. In Weyl semimetals, it is expected to induce a negative longitudinal magnetoresistance, the chiral magnetic effect.
Here, we present experimental evidence that the observation of the chiral magnetic effect can be hindered by
an effect called ``current jetting''. This effect also leads to a strong apparent negative longitudinal magnetoresistance, but it is characterized by a highly non-uniform current distribution inside the sample. It appears in materials possessing a large field-induced anisotropy of the resistivity tensor, such as almost compensated high-mobility semimetals due to the orbital effect. In case of a non-homogeneous current injection, the potential distribution  is strongly distorted in the sample. As a consequence, an experimentally measured potential difference is not proportional to the intrinsic resistance.
Our results on the magnetoresistance of the Weyl semimetal candidate materials NbP, NbAs, TaAs, and TaP exhibit distinct signatures of an inhomogeneous current distribution, such as a field-induced ``zero resistance'' and a strong dependence of the ``measured resistance'' on the position, shape, and type of the voltage and current contacts on the sample. A misalignment between the current and the magnetic-field directions can even induce a ``negative resistance''. Finite-element simulations of the potential distribution inside the sample, using typical resistance anisotropies, are in good agreement with the experimental findings. Our study demonstrates that great care must be taken before interpreting measurements of a negative longitudinal magnetoresistance as evidence for the chiral anomaly in putative Weyl semimetals.

%%%%%%%%%%%%%%%%%%%%%%%%%%%%%%%%%%%%%%%%%%%%%%%%%%%%%%%%%%%%%%%%%%%%%%%%%

\end{abstract}

%\pacs{74.25.Fy, 74.70.Dd}
%74.25.Fy Transport properties (electric and thermal conductivity, thermoelectric effects, etc.)
%74.20.Rp Pairing symmetries (other than s-wave)
%74.70.Dd Ternary, quaternary, and multinary compounds (including Chevrel phases, borocarbides, etc.)

\maketitle

%%%%%%%%%%%%%%%%%%%%%%%%%%%%      INTRODUCTION

\section{Introduction}

Weyl fermions are chiral massless fermions \cite{Weyl1929}. They can appear as quasiparticles in condensed-matter systems with strong spin-orbit coupling, where accidental band crossings cause a Dirac-like dispersion around discrete points in the Brillouin zone \cite{Wan2011, Liu:2014bf,Young_2012, Fang_2012}. In combination with a broken inversion or time-reversal symmetry, Dirac points split into pairs of Weyl points with opposite chirality. The term chirality denotes the fact that the Weyl points are either sources or drains of Berry flux. In the presence of parallel magnetic and electric fields an imbalance in the number of particles of each chirality  \cite{Adler1969,Bell1969} leads to the so-called chiral magnetic effect resulting in a negative longitudinal magnetoresistance (${\rm
MR}$) \cite{Nielsen1983}.

Recently, Weyl fermions were predicted near the Fermi energy in the non-centrosymmetric semimetals TaAs, TaP, NbAs, and NbP \cite{Weng:2014ue,Huang:2015uu}. Band-structure calculations were confirmed by quantum oscillation studies and angle-resolved photoemission spectroscopy, where linear band crossings and Fermi-arc surface states were discovered \cite{Liu:2014bf,Shekhar_2015, Klotz_2016,Arnold_2016_TaAs,Arnold_2016,Reis_2016,Lv2015TaAs,Xu2015TaAs,Yang2015TaAs,Xu_2015_SciAdv_TaP,Xu_2015_NatPhys,
Lv_2015_NatPhys_TaAs,Xu_2015_CPL_NbP,Xu_2016_NatCom_TaP,Liu_2016}. Furthermore, a negative longitudinal ${\rm MR}$ was detected for TaAs \cite{Huang_2015, Zhang_2016}, NbAs \cite{Yang_2015}, TaP \cite{Du_2016}, NbP \cite{ZWang_PRB_2016_NbP} and interpreted as the chiral magnetic effect. Similar results have been reported for the Dirac semimetals Na$_3$Bi \cite{Xiong_2015}, ZrTe$_5$ \cite{Zheng_2016}, and Cd$_2$As$_3$ \cite{HLi_2016}.  However, a negative longitudinal ${\rm MR}$ was also observed in materials, which where not expected to show a chiral magnetic effect, \textit{e.g.} in MPn$_2$ (M=Nb, Ta; Pn=As, Sb) \cite{YPLi_2016, Shen_2016, Luo_2016, YLi_2016}, in GdPtBi \cite{Hirschberger_2016}, or in the highly conductive delafossites PdCoO$_2$ and PtCoO$_2$ and in the ruthenate Sr$_2$RuO$_4$ \cite{Kikugawa_2016}. All these materials show a high charge carrier mobility which causes a magnetic field-induced resistance anisotropy  $A=\rho_{xx}/\rho_{zz}$. Due to the orbital effect in high-mobility materials, the resistivity normal to the magnetic field $\rho_{xx}$ increases significantly upon increasing field, whereas the resistivity parallel to the magnetic field $\rho_{zz}$ is not affected.

Other materials with a strong orbital ${\rm MR}$ are, for example, pure elemental bismuth, tungsten, or chromium \cite{Yoshida_1975, Collaudin_2015, Reed_1971}. In these materials, it is well known that an inhomogeneous current distribution can occur in magnetic fields, when the current is not injected homogeneously into the sample \cite{Yoshida_1975,Reed_1971, Ueda_1980}. For point-like current contacts and a large resistance anisotropy $A$ (induced by the magnetic field), the current flows predominantly parallel to the magnetic field direction, where the resistance is small. If the magnetic field $B$ is applied along the line connecting the current contacts, the electrical current forms a jet between them, hence the name ``current jetting'' \cite{Pippard}. In that case, the voltage probes on the surface of the sample decouple from the current and probe a smaller or even zero potential difference $V(B)$ \cite{Yoshida_1979_JAP_1}. As the resistance anisotropy grows upon increasing field, the measured voltage decreases leading to an apparent negative longitudinal ${\rm MR}$, defined here by ${\rm MR(B)} = V(B)/V_0$, $V_0= V(B = 0)$. Recently, indications of an inhomogeneous current distribution were reported in TaP \cite{Arnold_2016}.

In this paper, we provide evidence for a field-induced current-jetting effect in the Weyl semimetal candidates NbP, NbAs, TaP, and TaAs. We show that this effect can easily dominate the measurements of the longitudinal ${\rm MR}$. The apparent negative longitudinal ${\rm MR}$ induced by the current jetting is so strong that it might hide a potentially existing chiral magnetic effect in this class of materials. The inhomogeneous current distribution associated with it manifests itself in additional characteristics, such as a strong dependence of the ${\rm MR}$ on the contact geometry \cite{Yoshida_1976b_JPSJ} as well as a typical angular dependence of the ${\rm MR}$ as a function of the field direction \cite{Yoshida_1975,Yoshida_1976_JPSJ, Yoshida_1980_JAP_2}. Even a``negative resistance'' might be observed, if the sample is not \textit{perfectly} aligned with respect to the magnetic field \cite{Pippard}. All experimental findings are in good agreement with finite-element simulations of the potential distribution taking into account the actual sample and contact geometries as well as a field-induced resistance anisotropy originating only from the transverse orbital ${\rm MR}$. Our study indicates that measurements of the longitudinal ${\rm MR}$ in high mobility materials with field-induced resistance anisotropy have to be verified carefully in order to rule out any influence of an inhomogeneous current distribution on the data.

%%%%%%%%%%%%%%%%%%      METHODS

\section{Methods} \label{methods}

High-quality single crystals of NbP, NbAs, TaP, and TaAs were synthesized via a chemical vapor transport reaction. More details on the sample preparation and characterization can be found elsewhere \cite{Shekhar_2015, Arnold_2016}. For the electrical resistance measurements, bar-shaped samples were cut from large oriented single crystals. Typical dimensions of the samples were $0.4\times 0.3\times 2.0$~mm (width $\times$ thickness $\times$ length). The electrical current ($I=3$~mA) was applied along the long sample direction and the resistance was measured in a 4--probe geometry. We note that we define the current direction by the direction of the straight line connecting the two current electrodes. The contacts were made by spot welding platinum wires with a diameter of $25~\mu{\rm m}$ to the sample. The magneto-transport experiments were performed at $T=2$~K in magnetic fields up to $B=9$~T utilizing a Physical Property Measurement System (Quantum Design Inc.). The resistance data were collected in positive and negative fields and subsequently symmetrized in order to remove a Hall contribution present due to a small misalignment of the voltage contacts.  The residual resistivity of the samples used in this work was about $1.2~\mu\Omega$cm  (NbP), $7.3~\mu\Omega$cm (NbAs), $10.6~\mu\Omega$cm (TaP) and $21~\mu\Omega$cm (TaAs). Simulations of the potential distribution inside the samples were performed within a model based on current conservation using a finite element method, as implemented in the commercial software package COMSOL Multiphysics 5.2. For the simulations,  we used a constant electrical current of $I=3$~mA, which was injected in the front and end surfaces ($ac$ planes of the tetragonal crystal structure) near the top of the sample by point-like contacts.

\section{Results}

Among the Weyl semimetal candidate materials of the TaAs family, NbP is the one with the highest reported transverse magnetoresistance, 8\,500 at 1.85~K and 9~T, and an extremely large mobility of $5\times10^{6}{\rm cm^{2}V^{-1}s^{-1}} = 5\times10^2{\rm T^{-1}}$ \cite{Shekhar_2015}. This makes NbP best suited to study effects of the resistance anisotropy on the current distribution. In addition, in NbP the charge carriers at the Fermi surface are only comprised of topologically trivial electrons as its Weyl nodes lie 5 meV above the Fermi energy and chirality is ill-defined \cite{Klotz_2016}. Hence, a negative longitudinal $\rm MR$, cannot be ascribed to the chiral magnetic effect \cite{Nielsen1983}.
In the following we will discuss the observation of a negative longitudinal ${\rm MR}$ in NbP, NbAs, TaP, and TaAs. First, we will describe its main origin based on an inhomogeneous current distribution, as caused by a magnetic field-induced resistance anisotropy, on the example of NbP, followed by a discussion of NbAs, TaP, and TaAs. Finally, we will address further implications for the search of the chiral magnetic effect in Weyl semimetals.

\subsection{Transverse magnetoresistance}

%%%%%%%%%%%%%%%%%%%%%%%%%%%%%%%%%%%%%%%%%%%%%%%%%%%%%%%%%%%%%%%%%%%%%%%

\begin{figure}[t!]
\begin{center}
  \includegraphics[width=1.0\linewidth]{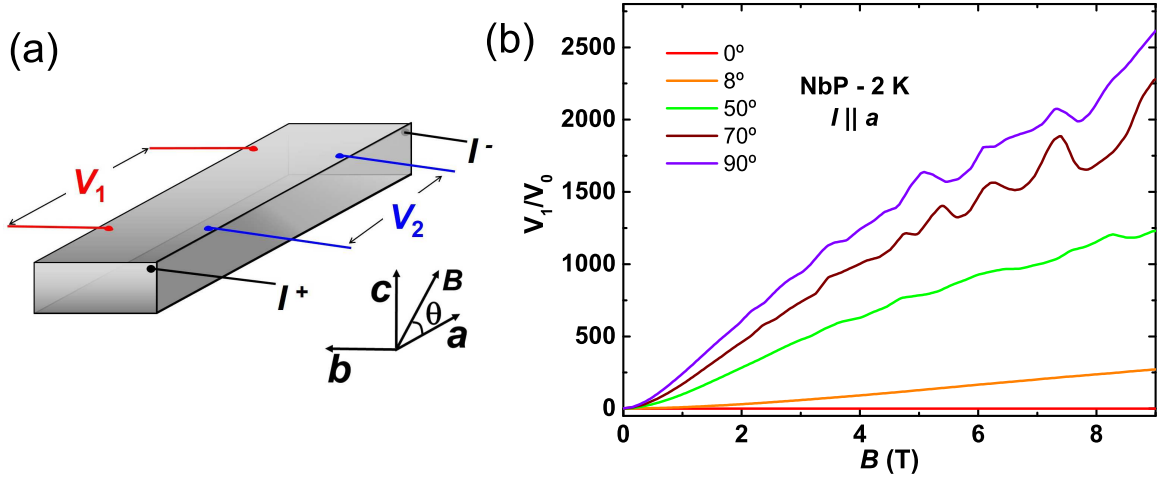}
  \caption{(a) Schematic drawing of the contact geometry used in our experiment. The $a$, $b$, and $c$ axes of the tetragonal crystal structure are in parallel to the sample edges as indicated. (b) Magnetoresistance of NbP at 2~K for different angles of the magnetic field $B$, as indicated. $V_1/V_0$ is the measured voltage for the contact pair $V_1$ normalized by the zero field value $V_0 = V(B = 0)$. The field is rotated in the $ac$ plane.
	}\label{MR}
  \end{center}
\end{figure}

%%%%%%%%%%%%%%%%%%%%%%%%%%%%%%%%%%%%%%%%%%%%%%%%%%%%%%%%%%%%%%%%%%%%%%%

To investigate the effect of the magnetic field on the current distribution in NbP, we set up a device where the electrical current is injected at the corner of the sample and two pairs of voltage contacts $V_1$ and $V_2$ are attached at different positions relative to the current electrodes. The geometry of our NbP sample with point-like electrodes is shown in Fig.~\ref{MR}a. In zero field, the values of $V_1$ and $V_2$ are identical within the experimental uncertainty when normalized by the contact separation. The magnetic field dependence of the MR is depicted in Fig.~\ref{MR}b for different angles $\theta$. Here, $\theta$ is defined as the angle between the magnetic field and the long sample axis ($a$) in the $ac$ plane. While we observe a different longitudinal ${\rm MR}(\theta=0)$ for the two voltage contacts pairs $V_1$ and $V_2$, which we will discuss in details below, the transverse ${\rm MR}(\theta=90^{\circ}, B \parallel c)$ are identical. The transverse ${\rm MR}$ increases by a factor of 2500 up to 9\,T at 1.85\,K and exhibits strong quantum oscillations. This demonstrates the high crystal quality and reflects the high mobility of the charge carriers. We note that our findings are in good agreement with the previously reported data \cite{Shekhar_2015}. The mobility estimated from the quadratic field dependence of the resistance at low fields is $2.8\times10^{2}{\rm T^{-1}}$. This is slightly smaller than the reported values in reference \cite{Shekhar_2015} in agreement with the slightly lower ${\rm MR}$.

\subsection{Longitudinal magnetoresistance}

%%%%%%%%%%%%%%%%%%%%%%%%%%%%%%%%%%%%%%%%%%%%%%%%%%%%%%%%%%%%%%%%%%%%%%%

\begin{figure}[t!b]
\begin{center}
  \includegraphics[width=0.8\linewidth]{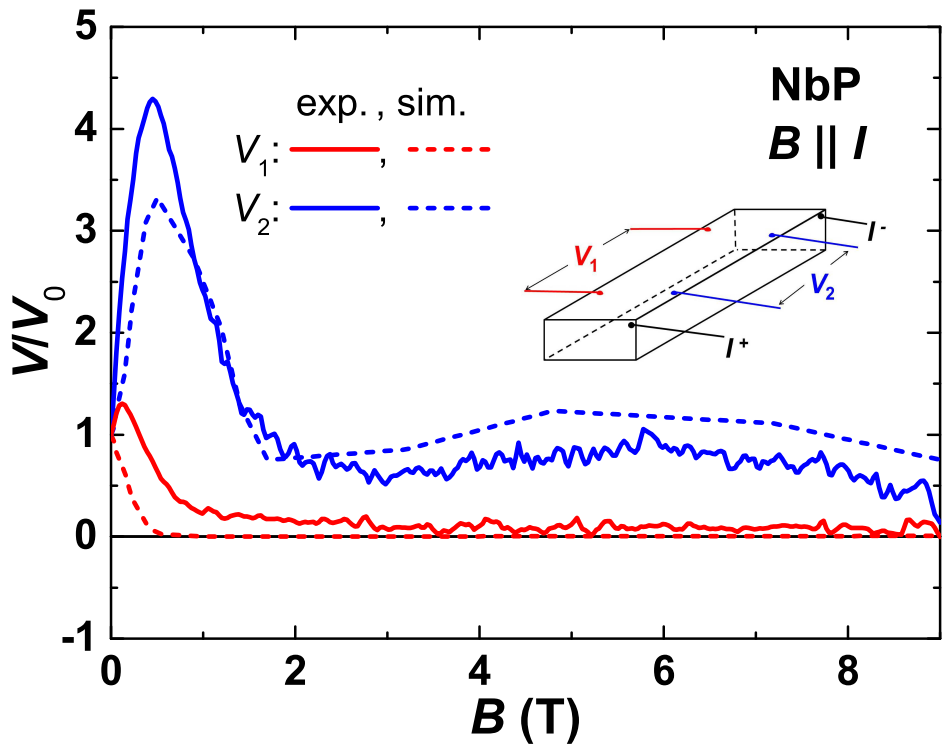}
  \caption{Experimental (solid lines) and simulated (dashed lines) magnetic field dependence of the voltage drop for two different contact pairs with the electrical current and magnetic field in parallel ($B\parallel I$). The fact that the two contact pairs exhibit a different field dependence reflects an inhomogeneous current distribution in the sample.
	}\label{Longitudinal_MR}
  \end{center}
\end{figure}

%%%%%%%%%%%%%%%%%%%%%%%%%%%%%%%%%%%%%%%%%%%%%%%%%%%%%%%%%%%%%%%%%%%%%%%

In the longitudinal configuration, where the magnetic field is applied parallel to the direction of the electrical current, $V_1(B)$ and $V_2(B)$ have a different magnetic-field dependence (see Fig.~\ref{Longitudinal_MR}). This observation is a direct hint at an inhomogeneous current distribution inside the sample induced by the magnetic field. Upon increasing the field $V_1(B)/V_0$ corresponding to the contact pair far away from the current electrodes first increases slightly, before the voltage drops to a very small value above 1\,T. On the other hand, for the voltage pair $V_2$, which is almost in line with the current contacts, the voltage initially increases much stronger than $V_1(B)/V_0$ until it reaches a maximum  at around 0.5~T, indicating a high current flow in this region of the sample at small fields. Upon further increasing $B$, $V_2(B)/V_0$ decreases rapidly displaying a minimum  at around 3\,T before a weak broad maximum appears. We note that for the same field $V_2(B)/V_0$ is always considerably larger than $V_1(B)/V_0$ (except for zero magnetic field). A similar field dependence of the longitudinal voltage, particularly the strong decrease upon increasing magnetic field, has been reported in chromium \cite{Reed_1971}, tungsten \cite{Reed_1971}, and bismuth \cite{Yoshida_1976b_JPSJ}, and was explained by the current jetting effect \cite{Pippard}. In order to get deeper insights into the influence of the magnetic field on the current distribution in NbP, we performed finite-element simulations of the potential distribution as  a function of the field-induced resistance anisotropy.

%%%%%%%%%%%%%%%%%%%%%%%%%%%%%%%%%%%%%%%%%%%%%%%%%%%%%%%%%%%%%%%%%%%%%%%

\begin{figure}[t!b]
\begin{center}
  \includegraphics[width=1.0\linewidth]{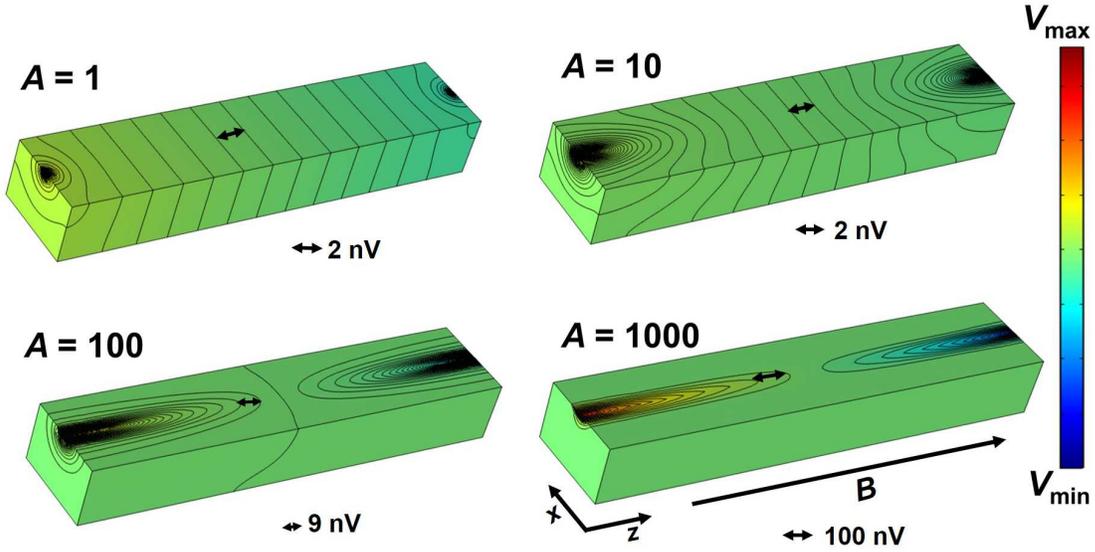}
  \caption{Calculated potential distribution for different resistance anisotropies $A= \rho_{xx}/\rho_{zz} =1$, 10, 100, and 1000 for a sample with point-like current electrodes and the dimensions of the investigated NbP sample. Note, the current contacts have been moved from the corner to the middle compared with the experimental set up for a better visualization. Here $B$ is aligned with the current electrodes, $\rho_{zz}$ ($\rho_{xx}$) is the resistance parallel (perpendicular) to $B$. The lines are contour lines of the equipotential surfaces. The increase in $\rho_{xx}$ and hence the anisotropy, induced by an increase in the magnetic field, strongly distorts the equipotential lines. }
\label{Simulation}
  \end{center}
\end{figure}

%%%%%%%%%%%%%%%%%%%%%%%%%%%%%%%%%%%%%%%%%%%%%%%%%%%%%%%%%%%%%%%%%%%%%%%

The potential distribution for various resistance anisotropies $A$ is depicted in Fig.~\ref{Simulation}. The sample has the same dimensions as the one studied experimentally, but the point-like current contacts have been attached in the middle of the front and end surfaces, close to the top of the sample for a better illustration. The magnetic field producing the resistance anisotropy is assumed to be parallel to the line connecting the current electrodes.
$A=1$ corresponds to the initial potential distribution for $\rho_{xx} = \rho_{zz}$ in zero magnetic field. The equipotential planes, except for a region close to the current contacts, are parallel indicating a homogeneous current distribution in most parts of the sample. Upon increasing the resistance anisotropy $A$ by increasing $\rho_{xx}$ and keeping $\rho_{zz}$ constant, we observe a dramatic effect on the potential distribution. Even for a small anisotropy of $A=10$, we find a pronounced distortion of the equipotential planes inside the sample leading to a strongly reduced region of a quasi-homogeneous potential distribution. For higher anisotropies, e.g.\ $A=100$ and $1000$, the equipotential planes are highly distorted as predicted for the current jetting effect \cite{Pippard}.

In order to extract the magnetic field dependence of the voltage from the simulations, we assume a field independent longitudinal resistance $\rho_{zz}(B) = \rho(B = 0)$. In that case the anisotropy is defined as $A(B) = \rho_{xx}(B \perp I)/\rho_0 = V_1(B, \theta=90^\circ)/V_0$ as given in Fig.~\ref{MR}. Note that for a magnetic field along the crystallographic $a$ axis, the transverse components of the resistance are generally not the same due to an anisotropy in the transverse MR in the crystallographic $bc$ plane. For our simulations, we neglect this anisotropy and assume  $\rho_{xx}=\rho_{yy}$. The simulations to determine $V_{1,2}(B)$ were carried out for the geometry of our sample and the actual positions of the contacts, in particular, the current was injected in the corners of the front and end surfaces in contrast to the visualization in Fig.~\ref{Simulation}. Hence, $A = 1$ corresponds to $B = 0$, $A = 10$ to 0.2\,T, $A = 100$ to 0.5\,T and $A = 1000$ to 3.2\,T. The calculated field dependence of the voltage for the two different configurations is shown as dashed lines in Fig.~\ref{Longitudinal_MR}. It reflects purely the effect of a field induced transverse resistance on the longitudinal measurements. The agreement between the simulations and the experiment is astonishing. Based on the results of our simulations, we argue that the pronounced voltage drop taking place upon increasing the magnetic field is caused by the intense increase of the transverse ${\rm MR}$ (and correspondingly of the anisotropy $A$), which forces the current to concentrate along a straight line connecting the current electrodes. Voltage electrodes on the edge of the sample far away from this line, such as $V_1$, detect a reduced voltage already for very small magnetic fields (small anisotropies). In our geometry even an anisotropy of only $A = 10$ ($\equiv 0.2$~T) is enough to reduce the voltage to 65~\% of the zero-field value. At the same time, the effective cross section of the sample is reduced, leading to higher currents and voltages close to the current jet. This is reflected in the pronounced increase of $V_2$ below 1~T. For higher fields (higher anisotropy $A$), the dilution even reaches the position of $V_2$. We point out that our simulations, based on a simple current conservation model, reproduce quantitatively all the main experimental features, namely the hump at low magnetic fields and the apparent negative longitudinal $\rm MR$. This suggests that the current jetting effect is the dominant cause of the experimentally observed \textit{negative} longitudinal $\rm MR$.

%%%%%%%%%%%%%%  Angular dependence
\subsection{Angular dependence of the magnetoresistance}

\label{AngDep}

%%%%%%%%%%%%%%%%%%%%%%%%%%%%%%%%%%%%%%%%%%%%%%%%%%%%%%%%%%%%%%%%%%%%%%%

\begin{figure}[t!b]
\begin{center}
  \includegraphics[width=1.0\linewidth]{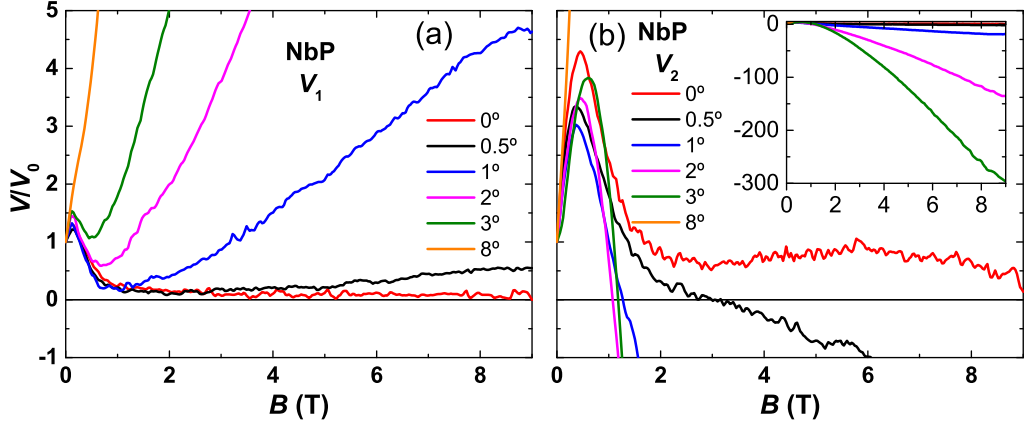}
  \caption{(a)  and (b) Magnetic field dependence of the voltage drop for different angles $\theta$ between the magnetic field and the electrical current for the two contact pairs $V_1$ and $V_2$, respectively. The inset in (b) displays the region where a negative voltage is observed for the contact pair $V_2$.
    }\label{Neg_Volt}
  \end{center}
\end{figure}

%%%%%%%%%%%%%%%%%%%%%%%%%%%%%%%%%%%%%%%%%%%%%%%%%%%%%%%%%%%%%%%%%%%%%%%

In the following, we will focus on the effect of a misalignment between the magnetic field and the current directions. In our experimental setup, the magnetic field is tilted away from the longitudinal configuration ($B \parallel I \parallel a$, $\theta = 0^{\circ}$) towards the transverse configuration ($B \parallel c \perp I$, $\theta = 90^{\circ}$) $(\theta$ is defined as in Fig.~\ref{MR}a). Special care was given to the alignment in the perpendicular direction, which is fixed on the rotator. Figure~\ref{Neg_Volt} illustrates the effect of a small misalignment on $V_1(B)$ and $V_2(B)$ for angles up to $\theta=8^{\circ}$. We point out the existence of strong characteristic differences in the field dependence of $V_1(B)$ and $V_2(B)$ in the small angle region. For $V_1(B)$ a very small angular misalignment of $\lesssim0.5\,^{\circ}$ is enough to obtain an increase in the voltage upon increasing the magnetic field (see Fig.~\ref{Neg_Volt}a). This is due to the projection of the transverse MR to the current direction. Moreover, for angles above $3^{\circ}$ the initial voltage drop present in the longitudinal configuration ($\theta=0^{\circ}$) starts to disappear. In the case of $V_2$ a misalignment leads to a characteristically different field dependence. Most strikingly, negative voltages are probed for misalignments between 0.5$^{\circ}$ and 3$^{\circ}$ in magnetic fields above 1\,T. A similar observation has been reported recently for the longitudinal $\rm MR$ of TaP, a member of the same family of materials \cite{Du_2016}. We will discuss the physical origin of the negative voltage values below.

%%%%%%%%%%%%%%%%%%%%%%%%%%%%%%%%%%%%%%%%%%%%%%%%%%%%%%%%%%%%%%%%%%%%%%%

\begin{figure}[t!b]
\begin{center}
  \includegraphics[width=1.0\linewidth]{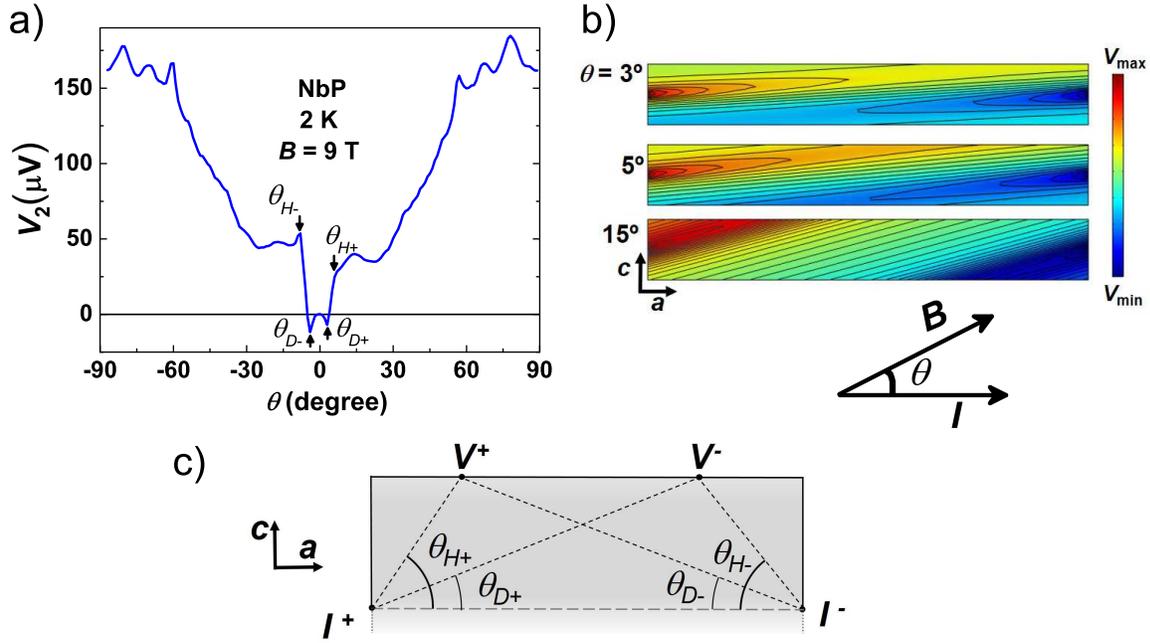}
  \caption{(a) Experimentally obtained angular dependence of the potential difference $V_{2}(\theta)$ for NbP in $B=9$~T. (b) Simulated potential distribution for $A = 2500$ (applied field of $B = 9$~T, $B\perp I$, in NbP) for  different angles $\theta$ in the $ac$ plane. This corresponds to a side view of the sample in Fig.~\ref{MR}. The current contacts have been attached on the front and end surfaces close to the middle of the edge parallel to $c$ for a better illustration. (c) Schematic illustration of the relation between the angles, where a dip ($\theta_D^+$, $\theta_D^-$) and a hump ($\theta_H^+$, $\theta_H^-$) appear in $V_2(\theta)$ and the geometry of the sample and positions of the electrodes.
  }\label{Angular_dependence}
  \end{center}
\end{figure}

%%%%%%%%%%%%%%%%%%%%%%%%%%%%%%%%%%%%%%%%%%%%%%%%%%%%%%%%%%%%%%%%%%%%%%%

The angular dependence of $V_2$ in a constant magnetic field of 9~T is displayed in Fig.~\ref{Angular_dependence}a. A negative voltage is detected, when the field direction is slightly tilted away from the longitudinal configuration. A dip appears in $V_2(\theta,B=9{\rm\, T})$ at an angle $\theta_D^+$ ($\theta_D^-$). Upon further increasing $\theta$, $V_2(\theta,B=9{\rm\, T})$ increases rapidly and a hump builds up at an angle $\theta_H^+$ ($\theta_H^-$). A similar angular dependence has been reported and explained for Bi and Sb samples in a similar geometry \cite{Yoshida_1975}. The strong variations in $V_2(\theta)$ near the transverse configuration ($\theta=90^{\circ}$) stem from quantum oscillations \cite{Shekhar_2015,Klotz_2016}.

As shown in Fig.~\ref{Simulation}, the equipotential lines are strongly distorted in the presence of a high magnetic field. The potential distribution from our simulation for different angles between the magnetic field and the current direction is shown in Fig.~\ref{Angular_dependence}b for $A = 2500$. As described before, the magnetic field is tilted in the $ac$ plane. Figures~\ref{Angular_dependence}b and c show a side view of the sample. For a better illustration, the current is injected in the middle of the side surfaces parallel to the $ac$ plane. Once the magnetic field is tilted away from the line connecting the current contacts, the equipotential lines start to tilt and to align parallel to the direction of the magnetic field \cite{Yoshida_1980_JAP_2}.

For high magnetic fields, corresponding to a large resistance anisotropy $A$, $\theta_D^{\pm}$ and $\theta_H^{\pm}$ are only determined by the geometry of the sample and the position of the electrodes as depicted in Fig.~\ref{Angular_dependence}c \cite{Yoshida_1976_JPSJ}. When the magnetic field is oriented parallel to the line connecting the positive current electrode $I^+$ and the closest voltage electrode $V^+$, this contact will be at a maximum of the potential and we observe a hump in the angular dependence ($\theta_H^+$  in Fig.~\ref{Angular_dependence}a). Similarly, when the magnetic field points towards the second voltage contact, the measured potential difference becomes negative and we observe a dip in the angular dependence ($\theta_D^+$ in Fig.~\ref{Angular_dependence}a). In our experiment, the agreement between the angles expected from the geometry,  giving $\theta_D^+(\theta_D^+$)$=2.6^{\circ} (-2.4^{\circ})$ and $\theta_H^+(\theta_H^-)=9^{\circ} (-7.5^{\circ})$ (marked by arrows in Fig.~\ref{Angular_dependence}a) and the position of the dips and humps in $V_2(\theta)$ is very good. The appearance and the height of the dip and the hump also depend strongly on the size of $A$, the thickness and the length of the sample, and the position of the voltage contacts with respect to the tilting plane of the magnetic field crossing the current electrodes \cite{Yoshida_1979_JAP_1, Yoshida_1976_JPSJ,Yoshida_1980_JAP_2}.
The dip and hump are most pronounced for voltage contacts lying in that plane (such as $V_2$) and disappear for voltage contacts far away (such as $V_1$, not shown here) \cite{Yoshida_1980_JAP_2}. The dips also vanish for very thin samples and become shallow for small $A$ \cite{Yoshida_1980_JAP_2}.

In conclusion, dips and humps can appear in the angular dependence of $V(\theta)$ characteristic of current jetting. For large resistance anisotropies, their position is only determined by the geometry of the sample and the position of the electrodes. This is a further evidence that the increase of the transverse resistance upon increasing field strongly influences any $\rm MR$ measurements in NbP and in similar materials. In general, for the longitudinal MR, the requirements on the alignment between the magnetic field and the current directions are extremely stringent (misalignment < 0.25 $^{\circ}$ in our setup). Otherwise, negative voltages can be observed.

%%%%%%%%%%%%%%%%%%       DISCUSSION

\section{Discussion}

The previous results, obtained on the example of NbP, are principally valid for all materials with a high mobility of the charge carriers and a field-induced anisotropy in the $\rm MR$. This in particular applies for the closely related materials of the same family: TaP, TaAs, and NbAs.

%%%%%%%%%%%%%%%%%%%%%%%%%%%%%%%%%%%%%%%%%%%%%%%%%%%%%%%%%%%%%%%%%%%%%%%

\begin{figure}[t!b]
\begin{center}
  \includegraphics[width=1.0\linewidth]{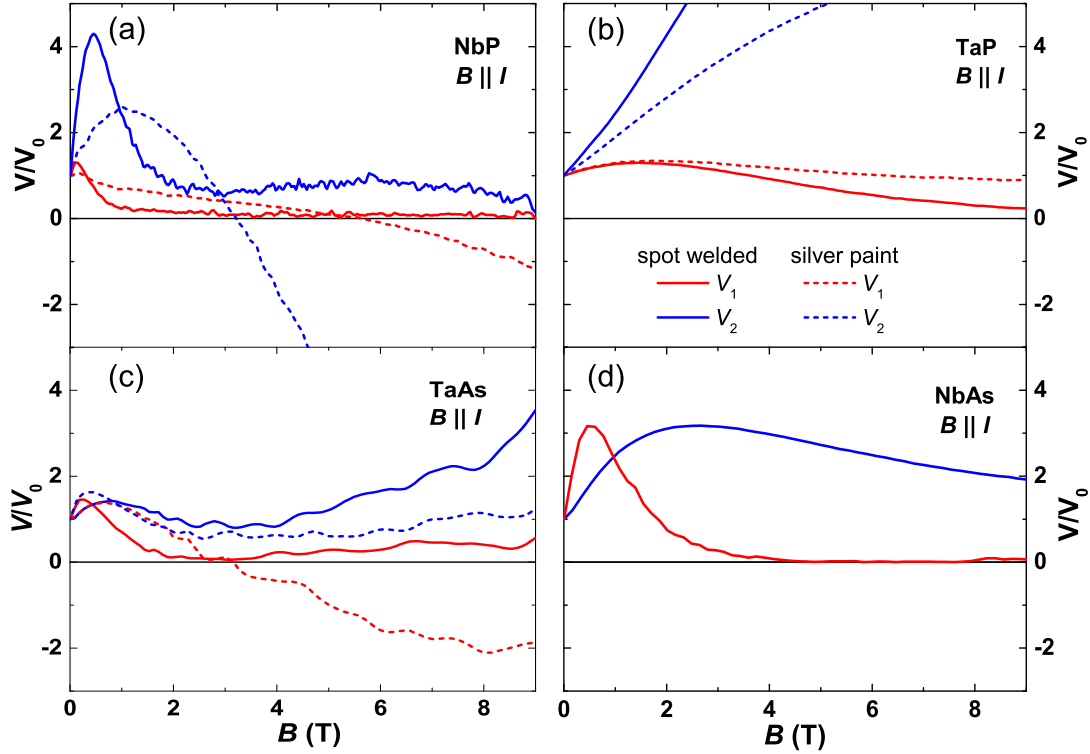}
  \caption{Field dependence of the voltage drop $V/V_0$  in (a) NbP, (b) TaP, (c) TaAs and (d) NbAs in the longitudinal configuration ($B\parallel I$). Two voltage contact pairs $V_1$ (red lines) and $V_2$ (blue lines) at different positions on the sample show a different field dependence due to an inhomogeneous current distribution.  In the first case (solid lines) point-like current electrodes were spot-welded to the samples and in the second case (dashed lines) the current electrodes were attached with silver paste covering the whole front and end surfaces of the samples. }
  \label{fig6}
  \end{center}
\end{figure}

%%%%%%%%%%%%%%%%%%%%%%%%%%%%%%%%%%%%%%%%%%%%%%%%%%%%%%%%%%%%%%%%%%%%%%%

Figure~\ref{fig6} displays the ``longitudinal $\rm{MR}$'' for the four members of the TaAs family measured with two different voltage contact pairs and two differently prepared pairs of current electrodes. Signatures of an inhomogeneous current distribution are present in all samples. The experimental setups with the spot-welded current contacts (solid lines) follow the schematic configuration illustrated in Fig.~\ref{MR}a. The corresponding $V(B)/V_0$ curves for the different voltage contact configurations $V_1$ and $V_2$ show a qualitative different field dependence. Here, $V_1$ is situated far from the line connecting the current electrodes and $V_2$ close to it. None of these curves reflects the intrinsic longitudinal $\rm MR$. Note that, in the case of TaAs, the increase of $V_1(B)$ and $V_2(B)$ in high fields points at a small misalignment between the current and the magnetic field directions (see Fig.~\ref{fig6}c).

However, the question remains, how the intrinsic longitudinal $\rm MR$ can be probed in order to extract the information if the chiral magnetic effect is present in these materials. There are two obvious experimental approaches improving the homogeneity of the current distribution, and a combination of both might give the best results.
The first is to inject the current homogeneously over the whole cross section of the sample. It turnes out that this is non-trivial. Soldering contacts is not possible in these materials and neither silver paste nor silver epoxy resulted in a satisfying outcome. The dashed lines in Fig.~\ref{fig6} mark the longitudinal voltages obtained with same voltage pairs as before, but using current electrodes made by silver paste applied over the whole front and end surfaces of the samples. There is still a significant difference between $V_1(B)$ and $V_2(B)$ for all materials. This is caused by a local variation of the contact resistance of the silver paint contacts leading to an injection of the current into the sample where the contact resistance is smallest. Since these positions are not necessarily the same on the front and end of the sample, the current direction is likely to be misaligned with respect to the direction of the magnetic field applied parallel to the sample. This might have been the case for experiments on NbP and TaAs (Fig.~\ref{fig6}a and c) where a negative voltage was detected (see also Sec.~\ref{AngDep}).

The second approach is to make the sample longer and/or thinner, so that the current has enough length to spread homogenously into the whole cross section of the sample. The critical parameter here is the aspect ratio $l/w$ of the sample length $l$ and width $w$ ($\rm width>thickness$). It is well known that the resistance anisotropy can be viewed as an effective shortening of the sample in the direction of the magnetic field by a factor of $1/\sqrt{A}$ (or an increase of $w$ by a factor of $\sqrt{A}$) \cite{Pippard}. Given a certain anisotropy of the resistivity $A$, a suitable ratio of $l/w$ for a homogeneous current distribution without resistance anisotropy ($A=1$) has to be be multiplied by a factor of $\sqrt{A}$ in order to still ensure a homogeneous current distribution. This is difficult to achieve for high $A$, which is determined in first approximation by the transverse MR. For example, in the investigated NbP sample with $A = 2500$, the required $l/w$ ratio would be higher than 50. Additionally, our simulations show that these long thin samples are much more sensitive to a misalignment of the magnetic field and electrical current directions. On the other hand, our experimental data and simulations show that an inhomogeneous current distribution appears already at very low fields (very small anisotropy $A$). Hence, for small ratios of $l/w$ ($ = 5$ for our sample of NbP) even a resistance anisotropy below 10 can be large enough to reduce significantly the region in the sample where the current is evenly distributed. As a result, an inhomogeneous current distribution can appear already at high temperatures, very low magnetic fields, and also in materials with low mobilities.

%%%%%%%%%%%%%%%%%%     CONCLUSION

\section{Conclusion}

We have studied the magnetoresistance of the putative Weyl semimetals of the TaAs family, where the chiral anomaly is expected to induce a negative longitudinal $\rm MR$. Indeed, an apparent negative $\rm MR$ is present in all investigated samples, NbP, NbAs, TaP, and TaAs. However, the field dependence of the \textit{measured} longitudinal $\rm MR$ is highly sensitive to the sample geometry and the position of the electrical contacts, in particular of the voltage contacts. Furthermore, it is extremely susceptible to the alignment between magnetic field and current directions. Small deviations from a parallel alignment even lead to an apparent \textit{negative} resistance.

Our experimental findings provide strong evidence for an inhomogeneous current distribution as the origin of the observed phenomena and seem to exclude the chiral magnetic effect as predominant cause. This inhomogeneous current distribution emanates from the field-induced resistance anisotropy in these high mobility materials and is known as current-jetting effect. Our conclusion is further supported by finite-element simulations based on a simple current-conservation model with the resistance anisotropy as only free parameter, which agree exceptionally well with the experimental results. An important implication of the simulations is that the intrinsic longitudinal magnetoresistance is almost constant in changing magnetic fields. We point out that the inhomogeneous current distribution already appears for very small resistance anisotropies below 10, which corresponds to small applied magnetic fields in our experiments. Moreover, our simulations expose that for high anisotropies it is very difficult to avoid signatures of the current-jetting effect even in samples with optimized geometry and ideal electrical contacts.

To conclude, our study demonstrates that measurement of the longitudinal $\rm MR$ in materials with field-induced resistance anisotropy is not straightforward and careful checking is required before intrinsic physical properties of the material, such as the chiral anomaly in Weyl semimetals, are extracted.

%%%%%%%%%%%%%%%%%  ACKNOWLEDGEMENTS

\section*{Acknowledgements}

We acknowledge stimulating discussions with J.\ Bardarson, K.\ Behnia, M.\ Brando, C.\ Felser, A.\ Grushin, B.\ Yan and especially K.\ Yoshida. We thank H. Borrmann for orienting the single crystals. R.\ dos Reis acknowledges financial support from the Brazilian agency CNPq (Brazil).

%%%%%%%%%% Bibliography %%%%%%%%%%%%%%%%%%%%%%%%%%%

%

\end{document}